\documentclass[a4paper,10pt]{article}
\usepackage[dvips]{graphicx}
\usepackage{amssymb,amsmath}
\numberwithin{equation}{section}

\begin{document}

\title{  G-essence cosmologies with  scalar-fermion  interactions}
\author{ Olga Razina$^1$,  Yerlan Myrzakulov$^1$, Nurzhan Serikbayev$^1$, Shynaray  Myrzakul$^1$,  \\ Gulgasyl Nugmanova$^1$  and   Ratbay Myrzakulov$^{1,2}$\footnote{The corresponding author. Email: rmyrzakulov@gmail.com; rmyrzakulov@csufresno.edu}\vspace{1cm}\\ \textit{$^1$Eurasian International Center for Theoretical Physics,} \\ \textit{Eurasian National University, Astana 010008, Kazakhstan}\\ \textit{$^2$Department of Physics, CSU Fresno, Fresno, CA 93740 USA} \\ \\ \\ \vspace{1cm}
 }

\date{}

\maketitle
\begin{abstract} We study the two particular models of g-essence  with Yukawa type 
interactions between  a scalar field $\phi$ and a classical  Dirac field $\psi$. For the  homogeneous, isotropic and flat Friedmann-Robertson-Walker universe 
 filled with the such g-essence, some  exact solutions of these models are found. 
 Moreover, we reconstruct the corresponding scalar and fermionic potentials.
\end{abstract}
\vspace{2cm} 

\sloppy

\section{Introduction} 
The recent data from type Ia supernovae and cosmic microwave background  radiation and so on \cite{Perlmutter}-\cite{Riess} have provided strong evidences for a spatially flat and accelerated expanding universe at the present time. This accelerated expansion of the universe is attributed to the domination of a component with negative pressure, called dark energy. So far, the nature of dark energy remains a mystery. In the literature, plenty of dark energy candidates has been proposed. Among of which are quintessence ($w>-1$), phantom ($w<-1)$, k-essence $(w>-1$ or $w<-1$), quintom ($w$ crosses $-1$), etc.  Note  that the accelerated expansion of the universe can also be obtained by the modified gravity theories like $F(R),\ F(G), \ F(T), \ F(R,T)$ and so on (see e.g. \cite{Momeni}). As is well-known,   the simplest model of dark energy  is the cosmological constant with energy density being near the vacuum energy $\rho_{\Lambda}\approx (10^{-3}eV)^{4}$ withoud varying with time. However, this  cosmological constant model of dark energy suffers from the severe problem of fine-tuning and coincidence. 

One of interesting models of dark energy is the \textit{k-essence}  \cite{Mukhanov1}-\cite{Mukhanov3} (see also \cite{Linder}-\cite{MR0}). Recently it is proposed the another model of dark energy - the so-called \textit{f-essence} \cite{MR1} which is the fermionic counterpart of the k-essence. Also it is proposed the \textit{g-essence} which is some hybrid construction of k-essence and f-essence (see e.g. \cite{MR1}-\cite{MR4}). 

 To our knowledge, in the literature there are relatively few works on the dark energy models with scalar and fermionic fields and with Yukawa type interactions (see e.g. refs. \cite{Zanusso}-\cite{Woodard2}).    In this paper, we study the g-essence cosmologies with Yukawa type interactions. 
 
 This paper is organized  as follows. In Section 2, we simply review the basic informations of g-essence in the Friedmann-Robertson-Walker (FRW)  universe. In Section 3, we present the particular model of g-essence with the Yukawa interactions and construct its exact solution. In Section 4, we introduce the more general particular model of g-essence and present its two exact  solutions. Section 5 is the conclusion  of the paper. 
 
\section{G-essence}

The action of     g-essence has the form 
\cite{MR1}
\begin {equation}
S=\int d^{4}x\sqrt{-g}[R+2K(X, Y, \phi, \psi, \bar{\psi})],
\end{equation} 
 where $K$ is some function of its arguments, $\phi$ is a scalar function, $\psi=(\psi_1, \psi_2, \psi_3, \psi_4)^{T}$  is a fermionic function  and $\bar{\psi}=\psi^+\gamma^0$ is its adjoint function. Here
\begin {equation}
X=0.5g^{\mu\nu}\nabla_{\mu}\phi\nabla_{\nu}\phi,\quad Y=0.5i[\bar{\psi}\Gamma^{\mu}D_{\mu}\psi-(D_{\mu}\bar{\psi})\Gamma^{\mu}\psi]
\end{equation}
are  the canonical kinetic terms for the scalar and fermionic fields, respectively. $\nabla_{\mu}$ and  $ D_{\mu}$ are the covariant derivatives. Note that the fermionic fields are treated here as classically commuting fields.  The model (2.1) admits important two reductions: \textit{k-essence} and \textit{f-essence}. 

We now  consider the  homogeneous, isotropic and flat FRW universe filled with g-essence. In this case, the metric  reads
	\begin{equation}
ds^2=dt^2-a^2(dx^2+dy^2+dz^2)
\end{equation}
and the vierbein is chosen to be (see e.g. \cite{Armendariz-Picon})
	\begin{equation}
	(e_a^\mu)=diag(1,1/a,1/a,1/a),\quad 
(e^a_\mu)=diag(1,a,a,a).
\end{equation}

 In the case of the FRW metric (2.3), the equations corresponding to the action (2.1) look  like \cite{MR1}
	\begin{eqnarray}
	3H^2-\rho&=&0,\\ 
		2\dot{H}+3H^2+p&=&0,\\
		K_{X}\ddot{\phi}+(\dot{K}_{X}+3HK_{X})\dot{\phi}-K_{\phi}&=&0,\\
		K_{Y}\dot{\psi}+0.5(3HK_{Y}+\dot{K}_{Y})\psi-i\gamma^0K_{\bar{\psi}}&=&0,\\ 
K_{Y}\dot{\bar{\psi}}+0.5(3HK_{Y}+\dot{K}_{Y})\bar{\psi}+iK_{\psi}\gamma^{0}&=&0,\\
	\dot{\rho}+3H(\rho+p)&=&0,
	\end{eqnarray} 
where  the kinetic terms, the energy density  and  the pressure  take the form
\begin{equation}
X=0.5\dot{\phi}^2,\quad  Y=0.5i(\bar{\psi}\gamma^{0}\dot{\psi}-\dot{\bar{\psi}}\gamma^{0}\psi)
  \end{equation}
  and 
\begin{equation}
\rho=2XK_{X}+YK_{Y}-K,\quad
p=K.
\end{equation}
\section{Model with Yukawa interactions} 
In this paper, we consider   the g-essence action (2.1) with
\begin {equation}
K=X+ Y- V_{1}(\phi)- V_2(u)-\eta \phi u,
\end{equation} 
where $\eta=const, \ u=\bar{\psi}\psi$. 	Then the  system  (2.5)-(2.10) takes the form
		\begin{eqnarray}
	3H^2-\rho&=&0,\\ 
		2\dot{H}+3H^2+p&=&0,\\
		\ddot{\phi}+3H\dot{\phi}+\eta u- V_{1\phi}=&0&,\\
			\dot{\psi}+\frac{3}{2}H\psi+i V^{'}_2 \gamma^0 \psi+i\eta\gamma^0\psi \phi&=&0,\\
		\dot{\bar{\psi}}+\frac{3}{2}H\bar{\psi}-i V^{'}_2\bar{\psi} \gamma^0 -i\eta\phi\bar{\psi}\gamma^0&=&0,\\
	\dot{\rho}+3H(\rho+p)&=&0,
	\end{eqnarray} 
	where
	\begin{eqnarray}
\rho&=&0.5\dot{\phi}^2+V_1+\eta \phi u+V_2, \\ 
p&=&0.5\dot{\phi}^2-V_1-V_2+V^{'}_2 u.
\end{eqnarray}

Let us now we show that the system (3.2)-(3.7) admits the exact analytical solutions. 
To do it, we will use the methods e.g. of \cite{Odintsov}-\cite{MR6}. In fact, for example, it has the following particular solution
\begin{eqnarray}
a&=&a_0t^{\lambda},\\ 
\phi&=&\phi_0t^{2-3\lambda},\\ 
\psi_l&=&\frac{c_{l}}{a^{1.5}_0t^{1.5\lambda}}e^{-iD},\quad (l=1,2),\\	
\psi_k&=&\frac{c_{k}}{a^{1.5}_0t^{1.5\lambda}}e^{iD}, \quad (k=3,4),
	\end{eqnarray}
where  $c_j$ obey the following condition $c=|c_{1}|^2+|c_{2}|^2|-|c_{3}|^2-|c_{4}|^2$,
\quad $\phi_0=-\frac{\eta c}{a^3_0 (2-3\lambda)}$
 and
 	\begin{equation}
D=-\frac{2\lambda a^3_0}{c(3\lambda-1)}t^{3\lambda-1}+\frac{\eta^2c}{3a^3_0(1-\lambda)}t^{3(1-\lambda)}
+D_{0}, \quad D_0=const.\end{equation}
The corresponding  potentials  take   the form
\begin{eqnarray}
V_1(\phi)&=&\frac{\phi^2_0\delta^2(\delta-1+3\lambda)}{2(\delta-1)}\left(\frac{\phi}{\phi_0}\right)^{\frac{2(\delta-1)}{\delta}}+\frac{\phi_0\delta\eta c}{a^3_0(\delta-3\lambda)}\left(\frac{\phi}{\phi_0}\right)^{\frac{\delta-3\lambda}{\delta}}+V_{10},\\ 
V_2(u)&=&3\lambda^2 \left(\frac{ua^3_0}{c}\right)^{\frac{2}{3\lambda}}+\frac{3\phi^2_0 \delta^2\lambda}{2(\delta-1)}\left(\frac{ua^3_0}{c}\right)^{-\frac{2(\delta-1)}{3\lambda}}+\frac{3\lambda\eta\phi_0 c}{a^3_0(\delta-3\lambda)}\left(\frac{ua^3_0}{c}\right)^{-\frac{\delta-3\lambda}{3\lambda}}-V_{10},
	\end{eqnarray}
where $V_{10}=const,\quad \delta=2-3\lambda,\quad u=\dfrac{c}{a^3}=\dfrac{c}{a_0^3 t^{3\lambda}}$.
For this solution, the equation of state and the deceleration parameters take the form
\begin{equation}
w=-1+\frac{2}{3\lambda}, \quad q=\frac{1-\lambda}{\lambda}.
\end{equation} So we conclude that,  in the case of the solution, 
the g-essence model (2.1) with (3.1) 
 can describes the accelerated expansion of the universe.

 \section{General model with Yukawa type interactions}

In this section, we consider   the g-essence action (2.1) with
\begin {equation}
K=\epsilon X+\sigma Y-V_1(\phi)-V_2(u)-\eta U_1(\phi)U_2(u),
\end{equation} 
where $\epsilon$ and $\sigma$ are some constants. Here we can note that  $\epsilon=1$
( $\epsilon=-1$)  corresponds to the usual (phantom) case. 
	Then the  system  (2.9)-(2.14) takes the form
		\begin{eqnarray}
	3H^2-\rho&=&0,\\ 
		2\dot{H}+3H^2+p&=&0,\\
		\epsilon\ddot{\phi}+3\epsilon H\dot{\phi}+\eta U_2 U_{1\phi}- V_{1\phi}&=&0,\\
			\sigma\dot{\psi}+\frac{3}{2}\sigma H\psi+i V_2^{'} \gamma^0\psi +i\eta U_1 U_2^{'}\gamma^0\psi&=&0,\\
		\sigma\dot{\overline{\psi}}+\frac{3}{2}\sigma H\overline{\psi}-i V_2^{'}\overline{\psi} \gamma^0 -i\eta U_1 U_2^{'}\overline{\psi}\gamma^0&=&0,\\
	\dot{\rho}+3H(\rho+p)&=&0,
	\end{eqnarray} 
	where
	\begin{eqnarray}
\rho&=&0.5\epsilon\dot{\phi}^2+V_1+\eta U_1 U_2+V_2, \\ 
p&=&0.5\epsilon\dot{\phi}^2-V_1-V_2+V^{'}_2 u.
\end{eqnarray}

As in the previous case,  we can construct the exact analytical solutions
of the system (4.2)-(4.7). 

i) As an example, here we present 
 the following particular solution
\begin{eqnarray}
a&=&a_0t^{\lambda},\\ 
\phi&=&\phi_0t^{\delta},\\ 
\psi_l&=&\frac{c_{l}}{a^{1.5}_0t^{1.5\lambda}}e^{-iD},\quad (l=1,2),\\	
\psi_k&=&\frac{c_{k}}{a^{1.5}_0t^{1.5\lambda}}e^{iD}, \quad (k=3,4),
	\end{eqnarray}
where  $c_j$ obey the following condition $c=|c_{1}|^2+|c_{2}|^2|-|c_{3}|^2-|c_{4}|^2$,
\quad $\phi_0=-\frac{\eta c}{a^3_0 (2-3\lambda)}$
 and
 	\begin{equation}
D=\frac{i}{\sigma}(-\frac{2\lambda a^3_0}{c(3\lambda-1)}t^{3\lambda-1}+\frac{\epsilon\phi^2_0\delta^2a^3_0}{c(2\delta-1+3\lambda)}t^{2\delta-1+3\lambda}+\frac{\eta a^3_0 U_{10} U_{20}}{c(n+l+3\lambda+1)}\left(1+\frac{n} {3\lambda}\right)t^{n+l+3\lambda+1}+D_0).\end{equation}
In this case, the corresponding  potentials  look like
\begin{eqnarray}
V_1(\phi)&=&\frac{\epsilon\phi^2_0\delta^2(\delta-1+3\lambda)(2\delta+1+3\lambda)}{2(\delta-1)(4(\delta-1)-n+3\lambda)} \left(\frac{\phi}{\phi_0}\right)^{\frac{2(\delta-1)}{\delta}}+V_{10},\\ 
V_2(u)&=&3\lambda^2 \left(\frac{ua^3_0}{c}\right)^{\frac{2}{3\lambda}}+\frac{3\epsilon\phi^2_0 \delta^2\lambda(2(\delta-1)-n-3\lambda)}{2(\delta-1)(4(\delta-1)-n+3\lambda)}\left(\frac{ua^3_0}{c}\right)^{-\frac{2(\delta-1)}{3\lambda}}-V_{10},\\
U_1(\phi)&=&-\frac{2\epsilon\phi^2_0\delta^2(\delta-1+3\lambda)}{\eta U_{20}(4(\delta-1)-n+3\lambda)}\left(\frac{\phi}{\phi_0}\right)^{\frac{2(\delta-1)-n}{\delta}},\\ 
U_2(u)&=&U_{20}\left(\frac{ua^3_0}{c}\right)^{-\frac{n}{3\lambda}},
	\end{eqnarray}
where $V_{10}=const,\quad \delta=2-3\lambda,\quad u=\dfrac{c}{a^3}=\dfrac{c}{a_0^3 t^{3\lambda}}$.
Also we can present the equation of state and the deceleration parameters for this solution. We have 
\begin{equation}
w=-1+\frac{2}{3\lambda}, \quad q=\frac{1-\lambda}{\lambda}.
\end{equation} So for  this solution, the g-essence model (2.1) with (4.1) 
 also can describes the accelerated expansion of the universe for some values of $\lambda$.
 
 ii) As the second, let us consider the de Sitter like solution. 
 It reads as
\begin{eqnarray}
a&=&a_0e^{\beta t},\\ 
\phi&=&\phi_0e^{\kappa t},\\ 
\psi_l&=&\frac{c_{l}}{a^{1.5}_0e^{1.5\beta t}}e^{-iD},\quad (l=1,2),\\	
\psi_k&=&\frac{c_{k}}{a^{1.5}_0e^{1.5\beta t}}e^{iD}, \quad (k=3,4),
	\end{eqnarray}
where  $c_j$ obey the following condition
 $c=|c_{1}|^2+|c_{2}|^2|-|c_{3}|^2-|c_{4}|^2$
 and
 	\begin{equation}
D=\frac{i}{\sigma}\left(\frac{\epsilon a^3_0  \phi^2_0 \kappa^2}{c(2\kappa+3\beta)}e^{(2\kappa+3\beta)t}+\frac{\eta a_0^3 U_{10} U_{20} }{c(l+n+3\beta)} \left(1+\frac{n}{3\beta}\right) e^{(l+n+3\beta)t}+D_{0}\right).\end{equation}
For the   potentials  we obtain the following expressions
\begin{eqnarray}
V_1(\phi)&=&\frac{\epsilon\phi_0^2 \kappa(\kappa+3\beta)(n+3\beta)}{2(4\kappa-n+3\beta)}\left(\frac{\phi}{\phi_0}\right)^{2}+V_{10},\\ 
V_2(u)&=&\frac{3\epsilon\phi^2_0\kappa\beta(2\kappa-n-3\beta)}{2(4\kappa-n+3\beta)}\left(\frac{ua^3_0}{c}\right)  ^{-\frac{2\kappa}{3\beta}}-V_{10}+3\beta^2,\\
U_1(\phi)&=&-\dfrac{2\epsilon\phi^2_0\kappa^2(\kappa+3\beta)}{\eta U_{20}(4\kappa-n+3\beta)}\left(\frac{\phi}{\phi_0}\right)^{\frac{2\kappa-n}{\kappa}},\\ 
U_2(u)&=&U_{20}\left(\frac{ua^3_0}{c}\right)^{-\frac{n}{3\lambda}},
	\end{eqnarray}
where $V_{10}, n=consts$. In this case,  the equation of state and the deceleration parameters take the form $w=q=-1$ that is we have the de Sitter spacetime.
  \section{Conclusion} 
In this paper, we have studied two particular cases of  g-essence with the Yukawa type interactions between the  scalar and the fermion fields. We constructed some examples of exact analytical solutions of these models.  The corresponding scalar and fermionic potentials are presented. These results   show that the g-essence with the Yukawa interactions can describes the accelerated expansions of the universe. Finally we would like to note that, quite recently,  the models with the classical fermionic fields were studied also in \cite{Kremer}-\cite{Chimento}. 
	\end{document}